\documentclass[aps,prb,amsmath,amssymb,twocolumn]{revtex4}
\usepackage{graphicx}
\usepackage{dcolumn}
\usepackage{bm}

\begin{document}

\title{Fulde-Ferrell-Larkin-Ovchinnikov state in disordered $s$-wave superconductors}

\author{Qinghong Cui and Kun Yang}

\affiliation{National High Magnetic Field Laboratory and Department of Physics, Florida State University, Tallahassee, Florida 32306, USA}

\date{\today}

\begin{abstract}
The Fulde-Ferrell-Larkin-Ovchinnikov (FFLO) state is a superconducting state stabilized by a large Zeeman splitting between up- and down-spin electrons in a singlet superconductor. In the absence of disorder, the superconducting order parameter has a periodic spatial structure, with periodicity determined by the Zeeman splitting. Using the Bogoliubov-de Gennes (BdG) approach, we investigate the spatial profiles of the order parameters of FFLO states in a two-dimensional $s$-wave superconductors with nonmagnetic impurities. The FFLO state is found to survive under moderate disorder strength, and the order parameter structure remains approximately periodic. The actual structure of the order parameter depends on not only the Zeeman field, but also the disorder strength and in particular the specific disorder configuration.
\end{abstract}

\maketitle

\section{Introduction}

In the early 1960's, Fulde and Ferrell,~\cite{FF1964} and independently Larkin and Ovchinnikov,~\cite{LO1965} proposed the possibility that superconducting states with periodic spatial variation of the superconducting order parameter would become stable, when a singlet superconductor is subject to a large Zeeman splitting. The Zeeman splitting could be due to either a strong external magnetic field, or an internal exchange field. Nowadays such states are collectively known as the Fulde-Ferrell-Larkin-Ovchinnikov (FFLO) state. Recently the FFLO state has received renewed interest due to experimental evidence of its existence in various superconductors,~\cite{gloos, tachiki, modler96a, modler96b, brooks, singleton, mazin, tanatar, annett, javorsky, radovan, bianchi03, capan, watanabe, martin, kakuyanagi, bianchi05, uji05, uji06, okazaki, young, correa, shinagawa, lortz, yonezawa} and the (thus far theoretical) possibility of its realization in trapped cold atom systems,~\cite{coldatom,yangreview} nuclear matter and in the core of neutron stars.~\cite{casalbuoni}

Soon after the original work of FFLO, it was found that the FFLO state is very sensitive to the presence of impurities, and eventually suppressed when the disorder strength reaches certain critical value.~\cite{aslamazov,takada,bulaevskii,agterbergyang} Physically, this is because the Cooper pairs in the FFLO state carry finite momenta (and the periodicity of the spatial structure is tied to the corresponding wavevectors); impurities cause scattering between states with different momenta and thus tend to destroy the corresponding spatial structure. In these earlier studies however, the effects of impurities are taken into account by disorder-averaging various relevant physical quantities (like pairing susceptibility); after such averaging translation symmetry is restored, and the superconducting order parameter is again assumed to have a periodic structure. On the other hand in a specific realization of disorder, the system will adjust itself according to the impurity configuration,~\cite{wang07} and its order parameter structure will no longer be exactly periodic. We can thus expect (possibly quite large) spatial fluctuations of the order parameter, which are neglected in these treatments. In fact such spatial fluctuations have been observed previously in strongly disordered superconductors in the {\em absence} of Zeeman splitting.~\cite{ma,ghosal98, moradian, lages, ghosal01} Thus a more complete understanding of the FFLO state requires a more careful examination of the disorder induced order parameter fluctuations. A related conceptual issue is how to distinguish the FFLO state from other competing states (including the Bardeen-Cooper-Schrieffer, or BCS state) in the presence of impurities, where momentum is no longer a good quantum number.

In this paper, we present a microscopic numerical study of a disordered, two-dimensional (2D) $s$-wave superconductor subject to a Zeeman field. We use a negative $U$ Hubbard model and treat it within the Hartree-Fock (HF) and Bogoliubov-de Gennes (BdG) framework. We solve for the order parameter configuration specific to each disorder configuration, and develop empirical criteria to distinguish among the BCS, FFLO and normal phases. A phase diagram based on such criteria is obtained.

This paper is organized as follows. In section II, we introduce the model Hamiltonian and present the mean-field treatment of our model. The numerical results of solutions of the order parameters are presented and analyzed in section III. Some concluding remarks are offered in section IV. Throughout this paper we only consider zero temperature, although generalization to finite temperature is straightforward.

\section{Model and mean-field treatment}

We begin with the disordered $s$-wave superconductor on a 2D square lattice of finite size $N_x \times N_y$, described by the negative $U$ Hubbard model:
\begin{align} \label{eq:hubbard}
\mathcal{H} =& -t \sum_{<\mathbf{i}, \mathbf{j}> \sigma} c_{\mathbf{i}\sigma}^{\dagger} c_{\mathbf{j}\sigma} + \sum_{\mathbf{i} \sigma} (w_{\mathbf{i}} + \sigma h - \mu) c_{\mathbf{i}\sigma}^{\dagger} c_{\mathbf{i}\sigma} \nonumber \\
& - U \sum_{\mathbf{i}} (\hat{n}_{\mathbf{i} \uparrow} - \frac{1}{2}) (\hat{n}_{\mathbf{i} \downarrow} - \frac{1}{2}),
\end{align}
where $t$ is the nearest-neighbor hopping that will be set $t=1$ from now on; $\sigma = \pm 1$ is the spin index; $w_i$ is the on-site random potential which is independently distributed from $-W/2$ to $W/2$ uniformly; $h$ is the Zeeman field; $U > 0$ is the attractive interaction; $\hat{n}_{\mathbf{i}\sigma}=c_{\mathbf{i}\sigma}^{\dagger} c_{\mathbf{i}\sigma}$ is the number operator, and $\mathbf{i} = (x_i, y_i)$ is the site position. Since the system size is finite, periodic boundary condition is imposed on the lattice. It is interesting to notice that the Hamiltonian (\ref{eq:hubbard}) is symmetric under the particle-hole transformation
\begin{equation}
c_{\mathbf{i}\sigma} = (-1)^{x_i + y_i} \tilde{c}_{\mathbf{i}\sigma}^{\dagger},
\end{equation}
in combination with the transformations: $\mu \to -\mu$, $h \to -h$ and $w_{\mathbf{i}} \to -w_{\mathbf{i}}$. This symmetry is respected by the mean-field approximation discussed below, and it is a very useful property that we use to test the accuracy of our numerical solutions.

Within the HF-BdG approximation, the Hamiltonian reduces to a quadratic form
\begin{align} \label{eq:mfhmlt}
\mathcal{H}_m =& \sum_{\mathbf{ij}\sigma} (H_{\mathbf{ij}} + \sigma h \delta_{\mathbf{ij}}) c_{\mathbf{i}\sigma}^{\dagger} c_{\mathbf{j}\sigma} + \sum_{\mathbf{i}} (\Delta_{\mathbf{i}} c_{\mathbf{i}\uparrow}^{\dagger} c_{\mathbf{i}\downarrow}^{\dagger} \nonumber \\
& + \Delta_{\mathbf{i}}^{\ast} c_{\mathbf{i}\downarrow} c_{\mathbf{i}\uparrow}),
\end{align}
where $H_{\mathbf{ij}} = -t \delta_{\mathbf{i} \pm \mathbf{1}, \mathbf{j}} + (w_{\mathbf{i}} - \mu - U \delta n_{\mathbf{i}\bar{\sigma}}) \delta_{\mathbf{ij}}$; $n_{\mathbf{i}\sigma} = \langle c_{\mathbf{i}\sigma}^{\dagger} c_{\mathbf{i}\sigma} \rangle$ is the particle density and $\delta n_{\mathbf{i}\sigma} \! = \! n_{\mathbf{i}\sigma} \! - \! 1/2$; $\bar{\sigma}=-\sigma$; $\Delta_{\mathbf{i}} = - U \langle c_{\mathbf{i}\downarrow} c_{\mathbf{i}\uparrow} \rangle$ is the order parameter. To diagonalize this HF-BdG (or mean-field) Hamiltonian (\ref{eq:mfhmlt}), we employ the Bogoliubov transformation
\begin{equation} \label{eq:btrans}
c_{\mathbf{i}\sigma} = \sum_{\nu} (u_{\mathbf{i}\sigma}^{\nu} \gamma_{\nu} - \sigma v_{\mathbf{i}\sigma}^{\nu\ast} \gamma_{\nu}^{\dagger})
\end{equation}
where $\gamma_{\nu}$ and $\gamma_{\nu}^{\dagger}$ are the quasiparticle operators. The amplitudes of the quasiparticles $(u_{\mathbf{i}\sigma}^{\nu}, v_{\mathbf{i}\sigma}^{\nu})$ satisfy the Bogoliubov-de Gennes (BdG) equations
\begin{equation} \label{eq:BdG0}
\sum_\mathbf{j} \! \left( \! \begin{array}{cc} H_{\mathbf{ij}\sigma} + h\delta_{\mathbf{ij}} & \Delta_{\mathbf{ij}} \\ \Delta_{\mathbf{ji}}^{\ast} & -H_{\mathbf{ij}\bar{\sigma}}^{\ast} + h\delta_{\mathbf{ij}} \end{array} \! \right) \!\! \left( \! \begin{array}{c} u_{\mathbf{j}\sigma}^{\nu} \\ v_{\mathbf{j}\bar{\sigma}}^{\nu} \end{array} \! \right) \! = \! E_{\nu} \!\! \left( \! \begin{array}{c} u_{\mathbf{i}\sigma}^{\nu} \\ v_{\mathbf{i}\bar{\sigma}}^{\nu} \end{array} \! \right),
\end{equation}
where the eigenvalues $E_{\nu} \ge 0$. The self-consistency conditions are consequently expressed as
\begin{subequations} \label{eq:self}
\begin{align}
& \Delta_{\mathbf{i}} = -U \sum_{\nu} [u_{\mathbf{i}\uparrow}^{\nu} v_{\mathbf{i}\downarrow}^{\nu\ast} f(E_{\nu}) - u_{\mathbf{i}\downarrow}^{\nu} v_{\mathbf{i}\uparrow}^{\nu\ast} f(-E_{\nu})], \\
& n_{\mathbf{i}\sigma} = \sum_{\nu} [|u_{\mathbf{i}\sigma}^{\nu}|^2 f(E_{\nu}) + |v_{\mathbf{i}\sigma}^{\nu}|^2 f(-E_{\nu})],
\end{align}
\end{subequations}
where $f(E)$ is the Fermi distribution function. Self-consistent solutions of $\Delta_{\mathbf{i}}$ and $n_{\mathbf{i}\sigma}$ are achieved by iteration of Eqs.~(\ref{eq:self}) from a randomly initialized configuration. In case different solutions are obtained from different initial configurations, we choose the energetically favored one by comparing their corresponding energies. The energy of the system in the superconducting state is evaluated to be
\begin{align} \label{eq:energy}
\langle \mathcal{H} \rangle =& \sum_{\nu} E_{\nu} [f(E_{\nu}) - \sum_{\mathbf{i}\sigma} |v_{\mathbf{i}\sigma}^{\nu}|^2] + \frac{1}{U} \sum_{\mathbf{i}} |\Delta_{\mathbf{i}}|^2 \nonumber \\
& + U \sum_{\mathbf{i}} n_{\mathbf{i}\uparrow} n_{\mathbf{i}\downarrow} - \frac{UN}{4}.
\end{align}
At zero temperature the Fermi function $f(E)$ reduces to a step function.

\section{Numerical Results}

In our numerical calculations, we take the chemical potential $\mu = -0.8$ to avoid the perfect nesting of the Fermi surface at half-filling ($\mu = 0$), and a relatively strong pairing interaction strength $U = 2.5$ so that the FFLO state can be realized in a relative small system size. We start by considering the disorder-free limit of $W=0$. In this case we can take advantage of the translation symmetry, and solve the problem in the thermodynamic limit. We find BCS order parameter $\Delta=0.365$ for $h=0$, and the FFLO state is stable for $h_{c1} < h < h_{c2}$ with $h_{c1} \approx 0.27$ and $h_{c2} \approx 0.66$. This field range is bigger than that of a 2D weak-coupling $s$-wave superconductor with circular Fermi surface.\cite{shimahara94} We believe the discrepancy is due to the fact that our pairing strength is not weak, and lattice effects that result in a non-circular Fermi surface with some nesting that favors the FFLO state. The system turns normal ($\Delta=0$) for $h > h_{c2}$. Throughout the range $h_{c1} < h < h_{c2}$, the FFLO order parameter takes a one-dimensional (1D) structure,~\cite{wang06} with a period that varies continuously with $h$.

In the presence of $W$, we will study systems with finite size of rectangular shape, which is fixed to be $N_x = 32$ and $N_y = 16$ for the rest of the paper. Periodic boundary condition is imposed in both directions. Here, we make the lattice geometry to be rectangular instead of square because this allows us to extend the size of one of the directions ($x$-direction) at the expense of the other, and the 1D structure of the FFLO state is not very sensitive to the size of the shorter direction. Having a finite systems size restricts the periodicity of the FFLO order parameter. At a given Zeeman field, if the lattice size is not commensurate with the corresponding period, solutions of the FFLO states will have to be modified to be commensurate with systems size, which results in some energy cost.

\begin{figure*}
\includegraphics[width = 0.9\textwidth]{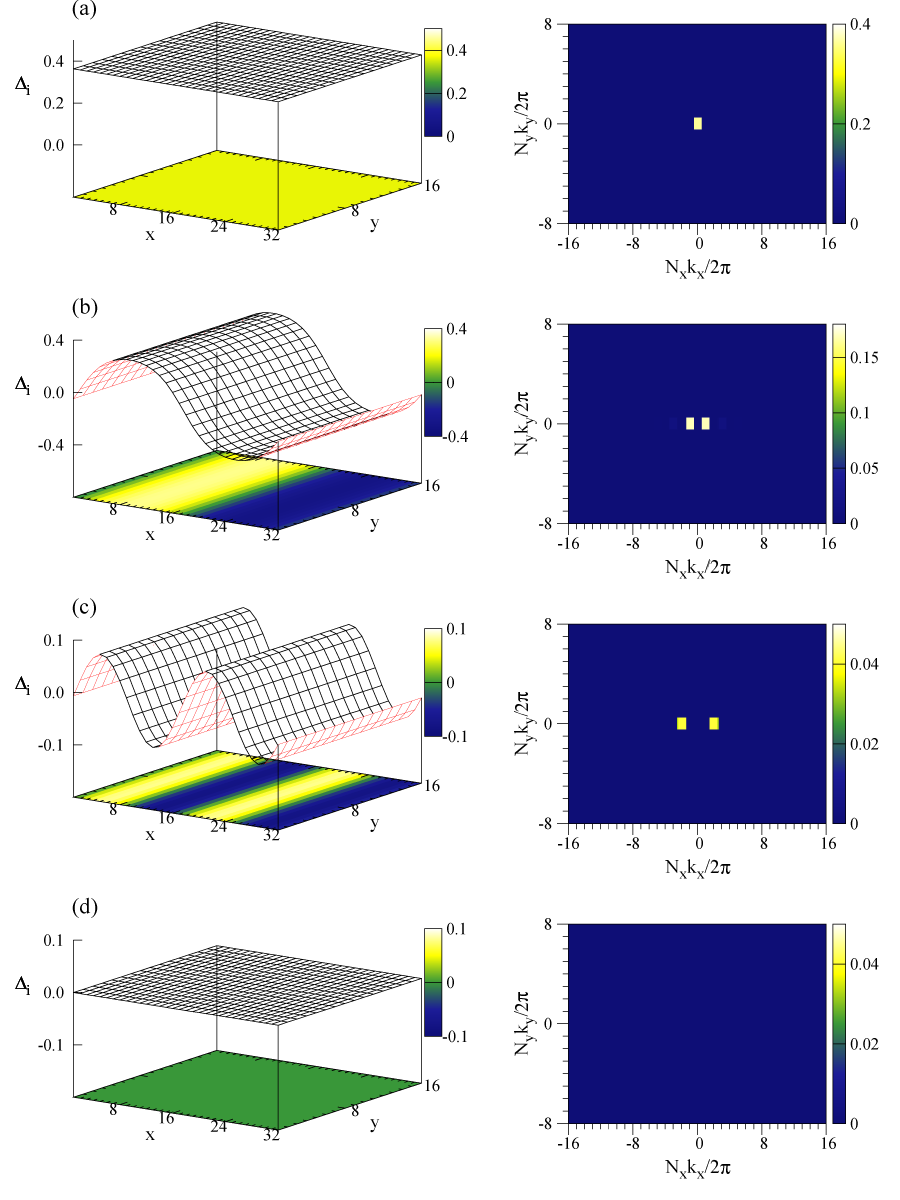}
\caption{\label{fig:op0} (color online) Order parameter at various $h$ in the absence of disorder ($W=0$) in both real space (left column) and momentum space (right column). (a) $h=0.0$, BCS state; (b) $h=0.27$, FFLO state with one period accommodated within the finite system size; (c) $h=0.35$, FFLO state with two periods accommodated within the finite system size; (d) $h=0.56$, normal state. Here, $U = 2.5$, $\mu = -0.8$, $N_x = 32$, $N_y = 16$.}
\end{figure*}

In the absence of disorder ($W=0$), FFLO state is obtained at this system size in the range from $h_{c1} = 0.27$ to $h_{c2} = 0.55$; the (slightly) reduced range reflects the finite size effect discussed above. The order parameters $\Delta_{\mathbf{i}}$ at several Zeeman fields are plotted in the left column of Fig.~\ref{fig:op0}, while the right column are the maps of the corresponding order parameters in Fourier space $\Delta_{\mathbf{k}}$,
\begin{equation}
\Delta_{\mathbf{k}} = \frac{1}{N_x N_y} \sum_{\mathbf{i}} \Delta_{\mathbf{i}} e^{-i \mathbf{k} \cdot \mathbf{x}_i}.
\end{equation}
Here, we see that the order parameter is of nearly cosine form with period of 32 along $x$-axis at $h =0.27$ [Fig.~\ref{fig:op0}(b)]. When the Zeeman field reaches $h=0.35$ [Fig.~\ref{fig:op0}(c)], the period reduces to 16 (so that two periods are accommodated), and it remains the same until the system becomes normal at $h=0.56$ [Fig.~\ref{fig:op0}(d)]. In the cases of the FFLO state, we find two dominant Fourier components representing the corresponding sinodal structure of the order parameter.

\begin{figure*}
\includegraphics[width = 0.9\textwidth]{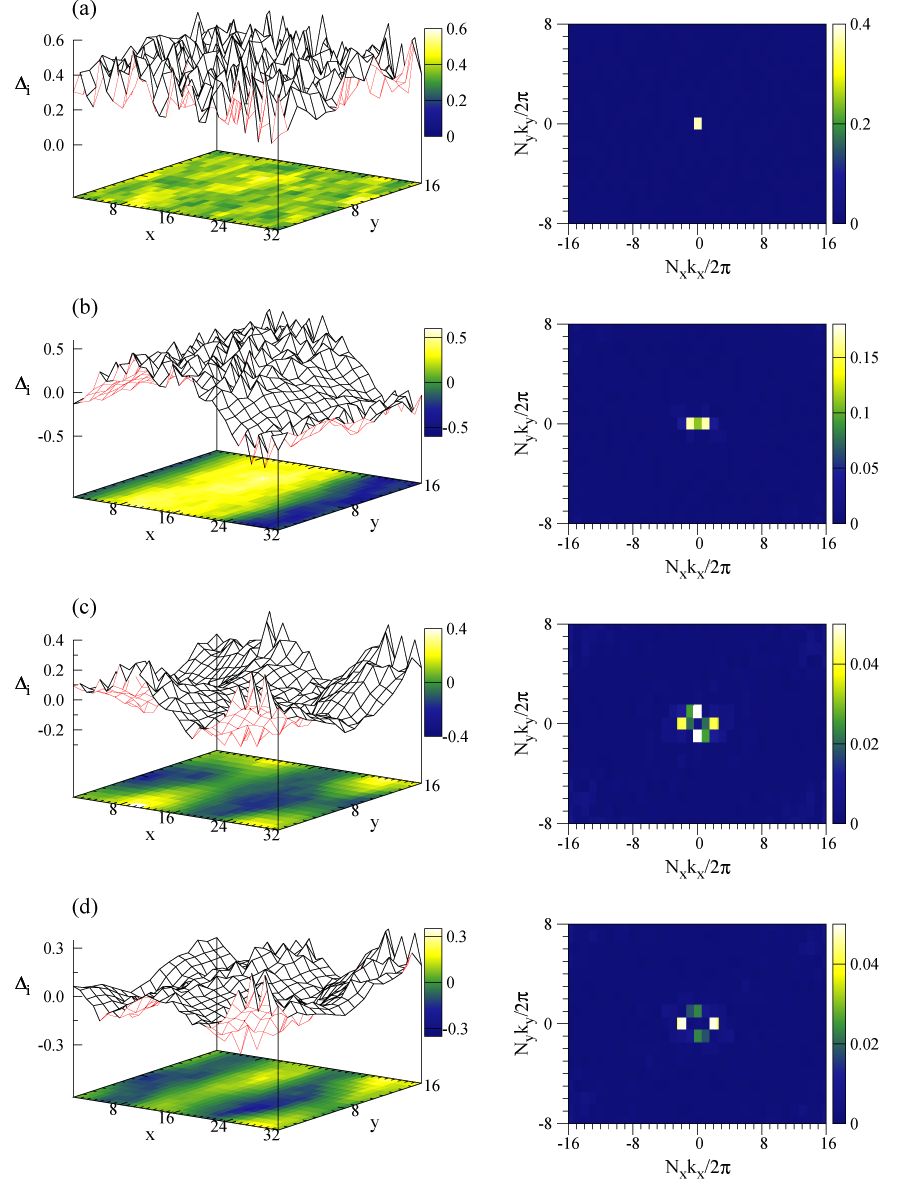}
\caption{\label{fig:op1} (color online) The same as Fig.~\ref{fig:op0} except $W=1.2$ and (a) $h=0.0$; (b) $h=0.32$, $A=0.299$, $B=-0.601$; (c) $h=0.36$, $A=0.019$, $B=-0.447$; (d) $h=0.40$, $A=0.012$, $B=-0.548$. See the text for the definitions of $A$ and $B$.}
\end{figure*}

\begin{figure*}
\includegraphics[width = 0.9\textwidth]{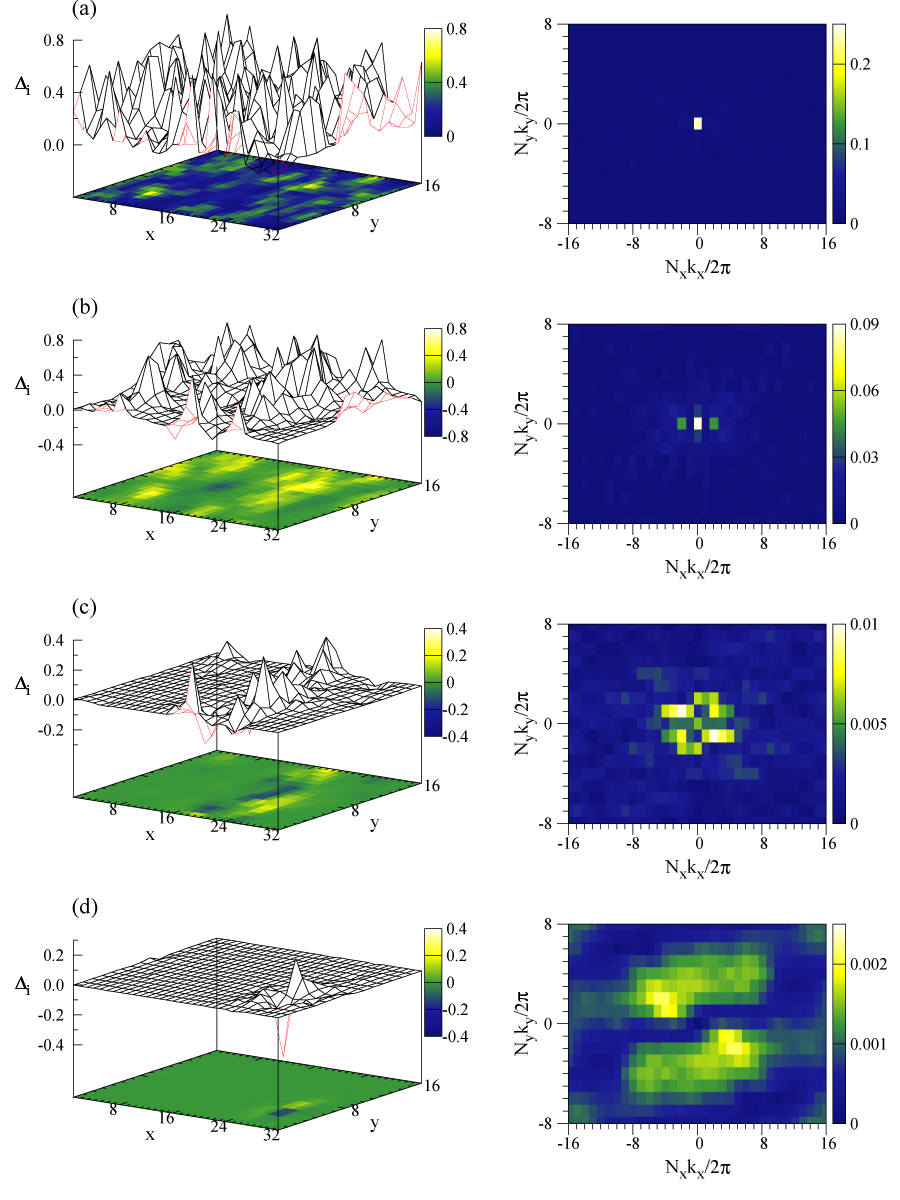}
\caption{\label{fig:op2} (color online) The same as Fig.~\ref{fig:op0} except $W=4.0$ and (a) $h=0.0$; (b) $h=0.34$, $A=0.411$, $B=0.511$; (c) $h=0.42$, $A=0.027$, $B=-0.195$; (d) $h=0.64$, $A=0.001$, $B=-0.099$. See the text for the definitions of $A$ and $B$.}
\end{figure*}

To obtain the physical picture of the $s$-wave superconductors subject to both disorder and Zeeman field, we calculate the order parameters $\Delta$ with $h = 0.24$--0.70 and $W = 0.0$--4.0. The disorder configurations are realized over 10 different samples. For comparison, we also calculate the order parameter $\Delta_0$ of each disorder configurations in the absence of Zeeman field, where Anderson's theorem guarantees a BCS state within mean-field theory. In Figs.~\ref{fig:op1} and \ref{fig:op2}, we give some examples of the order parameter configurations with various Zeeman fields under weak and strong disorder strengths ($W=1.2$ and $4.0$, respectively). Scrutinizing the spatial profiles of the order parameters individually leads to the following observations.

\begin{figure}
\includegraphics[width = 0.5\textwidth]{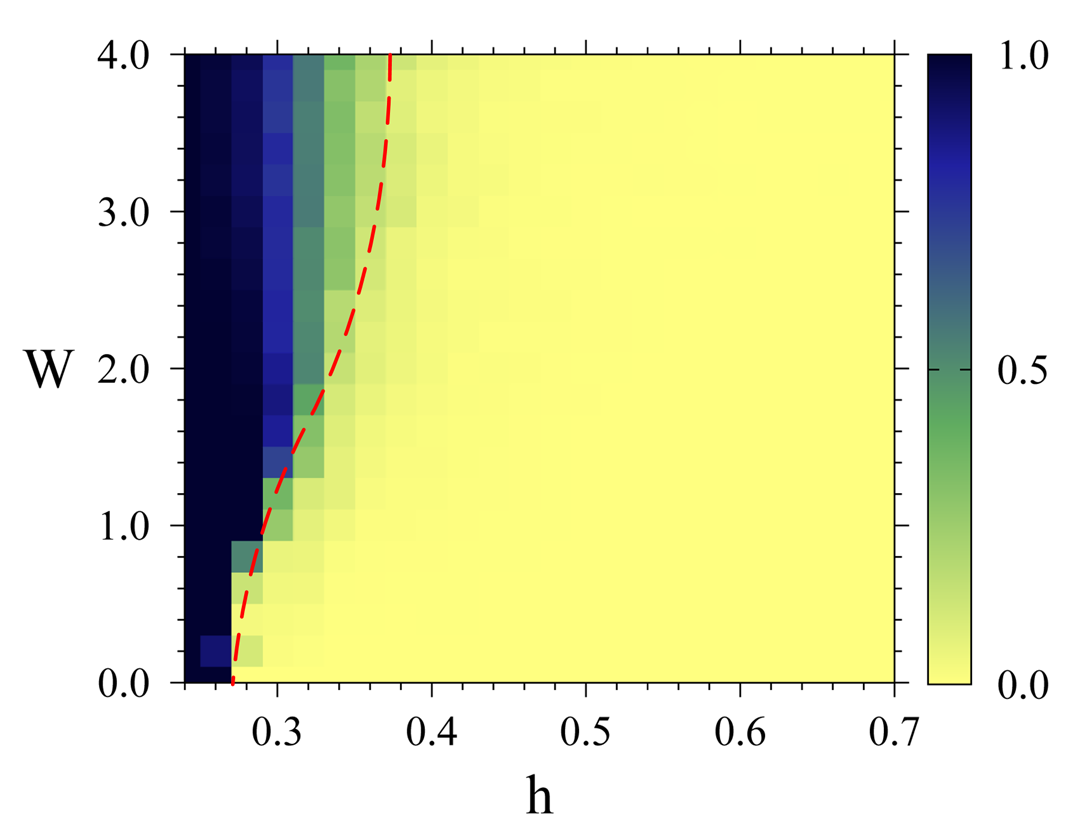}
\caption{\label{fig:phase1} (color online) The map of the overlap between $\Delta(h)$ and $\Delta_0(h=0)$, $A=|\langle \Delta_0| \Delta \rangle| / ||\Delta_0||^2$, as a function of Zeeman field $h$ and disorder strength $W$. We average over 10 samples of different disorder realizations for a given $W$. For the pure BCS state, the order parameter is unaffected by the Zeeman field and gives $A=1$. For non-BCS state, $A$ will decrease to zero rapidly. A rough boundary between the BCS state and non-BCS state is depicted by the (red) dashed line.}
\end{figure}

First of all, we find considerable sample specific spatial fluctuation in the order parameter in the BCS state [when $h=0$; see Figs.~\ref{fig:op1}(a) and \ref{fig:op2}(a)]. This is consistent with previous studies.~\cite{ghosal98, ghosal01} On the other hand the order parameter $\Delta$ has a single (positive) sign despite the fluctuations, as a result of which the Fourier transform of $\Delta$ peaks sharply at ${\bf k}=0$; this is a hallmark of the BCS state. When $h$ is turned on, there may be some sign changes in $\Delta$; but as long as $\Delta_{{\bf k}=0}$ remains the dominant Fourier component, we can still identify the state as BCS, and a typical example is Fig.~\ref{fig:op2}(b). Due to the presence of a gap, the BCS state is unaffected by the Zeeman field for a finite range of $h$. To determine the phase boundary across which the BCS state becomes unstable and yields to either the FFLO or normal state, we calculate the overlap between the order parameters in the presence and absence of Zeeman field,
\begin{equation}
A = |\langle \Delta_0 | \Delta \rangle| / ||\Delta_0||^2,
\end{equation}
where $||\Delta||=\sqrt{\langle \Delta|\Delta \rangle}$ is the norm of the order parameter. The reason we focus on this quantity is that the overlap between the order parameters of BCS state and either FFLO state or normal state is exactly zero for clean system, because the FFLO order parameter is oscillatory with zero mean, and the normal state has $\Delta=0$. We expect $A$ to decrease rapidly across the phase boundary where the BCS state destabilizes, and becomes almost zero in the FFLO and normal states with disorder. Numerical calculations of the sample averaged $A$ are displayed in Fig.~\ref{fig:phase1}. There we can see a fairly clear phase boundary separating the BCS and non-BCS states. The boundary is sharper in the weak disorder regime, while becomes somewhat fuzzy with increasing disorder strength due to larger sample-to-sample fluctuations.

We now turn to the more interesting FFLO state in the presence of disorder, examples of which are shown in Figs.~\ref{fig:op1}(b), \ref{fig:op1}(c) and \ref{fig:op1}(d). Just like the disorder-free case ($W=0$), the order parameters are oscillatory in these cases. But unlike the cases without disorder, the order parameters are no longer exactly periodic, and exhibit considerable sample specific spatial fluctuations similar to the BCS case with disorder. However the characteristic order parameter momenta are clearly visible in the Fourier space, as manifested by well defined peaks in $\Delta_{\bf k}$ {\em away from the origin} ${\bf k}=0$. Unlike the disorder-free cases however, these peaks have finite width, reflecting the absence of exact periodicity. We also find that disorder appears to influence the structure of the order parameter, such as in Figs.~\ref{fig:op1}(c) and \ref{fig:op1}(d), where the order parameters form 2D oscillatory structures while, with the same Zeeman field, only 1D structures appear in the absence of disorder. This is visible in real space but becomes particularly obvious in Fourier space, where $\Delta_{\bf k}$ has four instead of two peaks. We thus find that the order parameter structure depends not only on the Zeeman field, but also disorder strength and configuration.

Such disordered FFLO states are found at relatively weak disorder, up to $W\approx 2.0$. For stronger disorder a different pattern emerges as $h$ increases. In this case the BCS state yields to a state in which the order parameter is zero or nearly zero almost everywhere, but there can be isolated regions (or islands) within which $\Delta$ is non-zero and quite random. In Fourier space $\Delta_{\bf k}$ also shows a random form with no obvious structure, notably with $\Delta_{{\bf k}=0}\approx 0$. Examples of such states are Figs.~\ref{fig:op2}(c) and \ref{fig:op2}(d). This pattern is similar to what happens when increasing the disorder to very strong strength at $h=0$, where previous work found that the order parameter is destroyed everywhere except in a few superconducting islands.~\cite{ghosal98,ghosal01} The only difference here is that the order parameter tends to oscillate in sign in the presence of a Zeeman field. In Fourier space the order parameter shows random diffuse pattern with no clear structure in such states. Here we identify such states as the normal state, as in previous work.\cite{ghosal98,ghosal01} This is because while the order parameter is non-zero in isolated islands, quantum fluctuation will easily destroy the phase coherence between different islands, and as a result there is no phase coherence and stiffness in the thermodynamic limit.

It is clear from the discussions above that these three competing states (BCS, FFLO and normal) are most easily distinguished from the order parameter structures in Fourier space. To quantify their differences we introduce the  following quantity:
\begin{equation}
B = \eta |\Delta_{\mathbf{k}}|_{\mathrm{max}} / ||\Delta_{\mathbf{k}}||,
\end{equation}
where $\eta = +1$ if the maximum value of $|\Delta_{\mathbf{k}}|$ (denoted by $|\Delta_{\mathbf{k}}|_{\mathrm{max}}$) is at $\mathbf{k} = 0$ and $\eta = -1$ otherwise, and $||\Delta_{\mathbf{k}}||=\sqrt{\sum_{\bf k} |\Delta_{\mathbf{k}}|^2}$. In the absence of disorder, $B=1$ for the BCS state and $B=-1/\sqrt{2}$ for the FFLO state with 1D cosine structure (with only two Fourier components; the magnitude of $B$ reduces somewhat when more Fourier components are present). For the normal state, $B = 0$. In the presence of disorder, $\Delta_{\mathbf{k}}$ is dominated by the ${\bf k}=0$ component in the BCS state, and by several components with ${\bf k}\ne 0$ in the FFLO state; we thus expect B to be positive for the BCS state, and negative for the FFLO state. For the normal state on the other hand, there will be no dominant component in $\Delta_{\mathbf{k}}$; as a result $B$ is nearly zero for that case. To illustrate this identification we have listed the values of $B$ for the individual cases in Figs.~\ref{fig:op1} and \ref{fig:op2}.

\begin{figure}
\includegraphics[width = 0.5\textwidth]{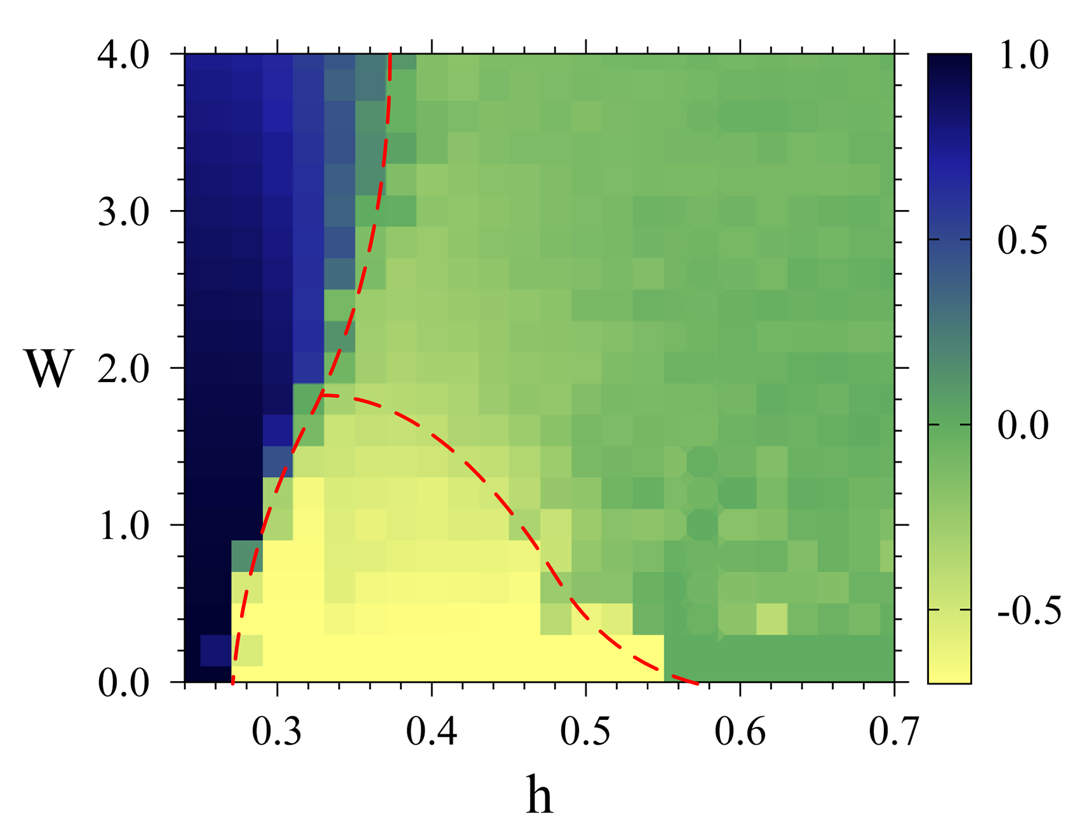}
\caption{\label{fig:phase2} (color online) The map of $B=\eta |\Delta_{\mathbf{k}}|_{\mathrm{max}} / ||\Delta_{\mathbf{k}}||$ as a function of Zeeman field $h$ and disorder strength $W$, where $\Delta_{\mathbf{k}}$ is the Fourier transform of $\Delta_{\mathbf{i}}$, $\eta$ takes the value of 1 when the maximum value $|\Delta_{\mathbf{k}}|_{\mathrm{max}}$ is at $\mathbf{k} = 0$, and of $-1$ otherwise. The disorder realizations are the same as in Fig.~\ref{fig:phase1}. In the absence of disorder, this quantity equals to 1 for the BCS state, $-1/\sqrt{2}$ for the FFLO state of 1D structure while $-1/2$ for 2D structure with simple cosine functions for the order parameter, and zero for the normal state. The (red) dashed lines sketch the (rough) phase boundaries among these three phases with disorder.}
\end{figure}

\begin{figure}
\includegraphics[width = 0.38\textwidth]{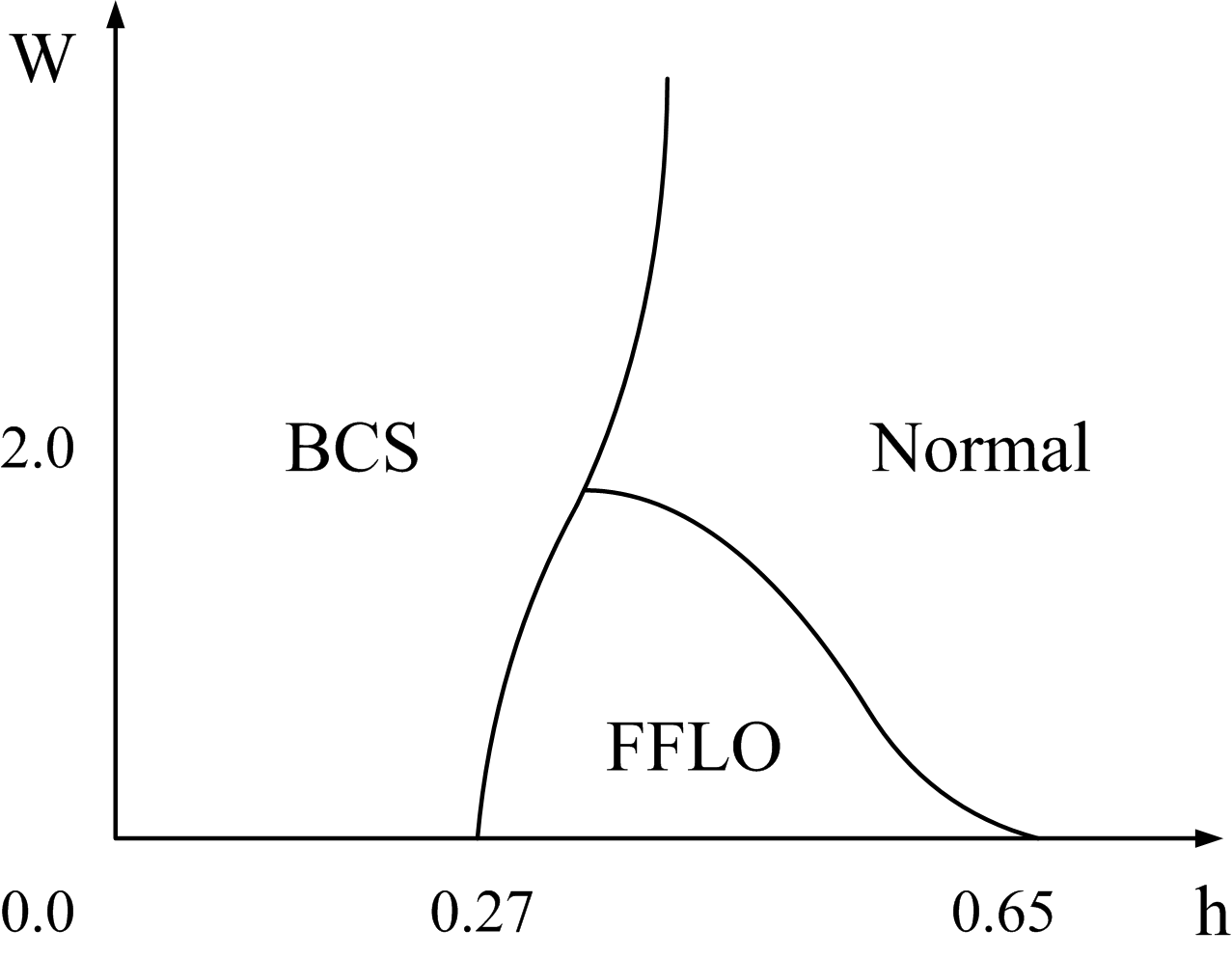}
\caption{\label{fig:phase3} Sketch of the phase diagram of disordered $s$-wave superconductors under Zeeman field, based on data presented in Figs.~\ref{fig:phase1} and \ref{fig:phase2}. The two critical Zeeman fields at $W=0$ are based on our numerical results of the same $U$ and $\mu$ as in Figs.~\ref{fig:phase1} and \ref{fig:phase2}, except that the lattice size is allowed to vary.}
\end{figure}

The disorder averaged $B$ value is plotted in Fig.~\ref{fig:phase2}. Using the identification above, we can clearly see these three phases, and identify three phase boundaries: BCS to FFLO, BCS to Normal, and FFLO to normal. The first two boundaries agree with the BCS to non-BCS boundary obtained from Fig.~\ref{fig:phase1}. By combining data in these two figures as well as our knowledge of the positions of $h_{c1}$ and $h_{c2}$ in the absence of disorder, we obtain a phase diagram of the system as presented in Fig.~{\ref{fig:phase3}.

\section{Concluding Remarks}

In this paper we have presented a microscopic study (based on mean-field theory) of a 2D $s$-wave superconductor subject to both disorder potential and a Zeeman field, and showed that the FFLO state can survive moderate disorder strength. Our most important results are presented in the form of a zero temperature disorder-Zeeman field phase diagram (Fig.~\ref{fig:phase3}), as well as the detailed analysis of the superconducting order parameters in different phases that leads to the determination of this phase diagram (Figs.~\ref{fig:phase1} and \ref{fig:phase2}). We should note that within mean-field theory we always obtain a BCS superconducting solution for zero Zeeman field ($h=0$); however quantum fluctuations that are left out in our study gets enhanced with increasing disorder strength $W$, and will eventually drive the system to an Anderson insulating phase. Thus the BCS-Normal phase boundary will eventually bend left as $W$ increases, and intersect the $h=0$ axis. Since we do not expect metallic behavior in 2D, the entire normal phase is expected to be an Anderson insulator. We also note that the superconducting islands we found in the ``normal" phase (which is really insulating) have also been observed in numerical studies of superconductor-insulator transition in disordered superconductors.\cite{dubi}

Our results are in qualitative agreement with previous studies\cite{aslamazov,takada} which suggested that the the FFLO state can survive as long as the electron scattering rate is comparable or smaller than the BCS gap. In our study we find the critical disorder strength $W_c \approx 2.0$; this corresponds to a scattering rate (within effective mass approximation) $1/\tau_c\approx W_c^2/(24\hbar t)\approx 0.167$ (we set $\hbar=1$), which is indeed comparable the the BCS gap $\Delta=0.365$. More importantly however, our study reveals that the FFLO state is characterized by {\em approximate}, instead of exact periodic variation of the superconducting order parameter in real space with {\em zero} average; this is manifested in Fourier space by peaks in the order parameter at non-zero wavevectors with finite width.

In an earlier work,~\cite{kun2000} one of us (along with D. F. Agterberg) proposed to use the Josephson effect between a BCS superconductor and an FFLO superconductor to identify the latter and measure its order parameter structure. It was shown that~\cite{kun2000} there is no Josephson effect in this case when there is no magnetic flux going through the junction, because there is no ${\bf k}=0$ component of the order parameter in the FFLO state. This conclusion remains to be true in the presence of disorder. Ref.~\onlinecite{kun2000} further predicts that the Josephson effect can be recovered when there is an appropriate amount of magnetic flux going through the junction, which gives rise to a momentum (or wavevector) to the Josephson hopping matrix element that matches momentum of one of the Fourier components of the FFLO order parameter; this allows for a direct measurement of the order parameter momenta in the absence of disorder. In the presence of disorder this method should still work as long as the disorder strength is sufficiently weak and the size of the junction is relatively small. However for stronger disorder and/or very large junction this will no longer work because the weight carried by a single Fourier component of the order parameter will be negligible in the disordered FFLO state.

It would be highly desirable to go beyond mean-field theory and take into account effects of quantum fluctuations of the order parameter. This is perhaps most easily done in one- or quasi-one-dimensional systems, where the powerful method based on bosonization has already yielded a number of exact results in the absence of disorder.~\cite{yang01} Disorder can be treated using the formalism advanced by Giamarchy and Schulz.~\cite{gs} We note that in 2- or 3-dimensional systems, initial attempts to go beyond mean-field theory have been made by treating fluctuations within RPA\cite{yokoyama} and fluctuation exchange\cite{yanase} approximations.

Among existing candidates of FFLO superconductors, CeCoIn$_5$ appears to be the most promising.~\cite{radovan,bianchi03} This is a $d$-wave superconductor. The $d$-wave FFLO state is different from $s$-wave FFLO state in a number of important ways.~\cite{maki1996,Kun1998,agterbergyang} It would thus be very interesting to use the methods of the current paper to study $d$-wave supercondutors in the presence of both disorder and Zeeman field, and compare with existing theory.

\acknowledgements
The authors thank Prof.~C.-R.~Hu for helpful discussions. This work was supported by National Science Foundation grant No.~DMR-0704133 (KY), NNSA/DOE grant DE-FG52-06NA26193 (QC), and the State of Florida (QC).

\bibliographystyle{prsty}

\end{document}